\documentclass[sn-mathphys,Numbered]{sn-jnl}

\usepackage{graphicx}%
\usepackage{multirow}%
\usepackage{amsmath,amssymb,amsfonts}%
\usepackage{amsthm}%
\usepackage{mathrsfs}%
\usepackage[title]{appendix}%
\usepackage{xcolor}%
\usepackage{textcomp}%
\usepackage{manyfoot}%
\usepackage{booktabs}%
\usepackage{algorithm}%
\usepackage{algorithmicx}%
\usepackage{algpseudocode}%
\usepackage{listings}%

\begin{document}

\title[Liquid $^3$He Detectors for Neutrons]{Liquid $^3$He Detectors for Neutrons}


\author*[1]{\fnm{A.N.} \sur{Villano}}\email{anthony.villano@ucdenver.edu}



\affil*[1]{\orgdiv{Department of Physics}, \orgname{University of Colorado Denver}, \orgaddress{\street{1200 Larimer St.}, \city{Denver}, \postcode{80217}, \state{Colorado}, \country{USA}}}




\abstract{
The $^3$He(n,p) process is excellent for neutron detection between thermal and $\sim$4\,MeV
because of the high cross section and near-complete energy transfer from the neutron to the
proton. Traditional gaseous $^3$He detectors using this process typically have high levels of
radiogenic backgrounds so that they cannot measure the small neutron fluxes present in underground
labs for dark matter experiments. I propose a cryogenic liquid $^3$He detector that can be
designed with tiny radiogenic backgrounds and efficiently measure neutron fluxes in low-flux
environments.

}




\maketitle

%
%
%

%
%
%
%

\section{\label{sec:intro}Introduction}

Low background underground environments, like SNOLAB, are home to many dark matter searches due to
their minimization of non-signal backgrounds. The underground environments typically mitigate
neutrons from cosmic rays (cosmogenic) very well, but there remains a neutron flux in the
environment, primarily due to radiogenic sources--spontaneous fission or ($\alpha$,n) from the
surrounding material~\cite{COOLEY2018110}. The levels of these backgrounds are so low that they
are difficult to measure, but still pose a threat to dark matter detectors as a background. My
work aims to develop a prototype liquid $^3$He scintillation detector to measure the neutron
backgrounds in low flux environments.  The high cross section reaction $^3$He(n,p)$^3$H in
combination with the high $^3$He number density in the liquid state will allow these low fluxes to
be measured by this prototype~\cite{sharbaugh2023response}. Measuring energy deposits as
low as 500~\,eV in liquid $^3$He has been previously demonstrated~\cite{Bradley1995} including
work from the ULTIMA collaboration explicitly searching for low-energy recoils from dark
matter~\cite{WINKELMANN2006384}.  Experiments have also specifically observed the $^3$He(n,p)
process in liquid $^3$He~\cite{Bauerle1996,Ruutu1996}. The scintillation of $^3$He should be
similar to that of $^4$He and scintillation yield has been measured for liquid
$^4$He~\cite{PhysRevD.105.092005} and reported in~\cite{sharbaugh2023response}. In the neutron
energy range below $\sim$5\,MeV the product proton travels less than 0.1\,mm and thus very nearly
all of the stopping energy of the proton and triton is deposited in the liquid volume.

\begin{figure}[h!]
    \centering
    \includegraphics[width=1.0\linewidth]{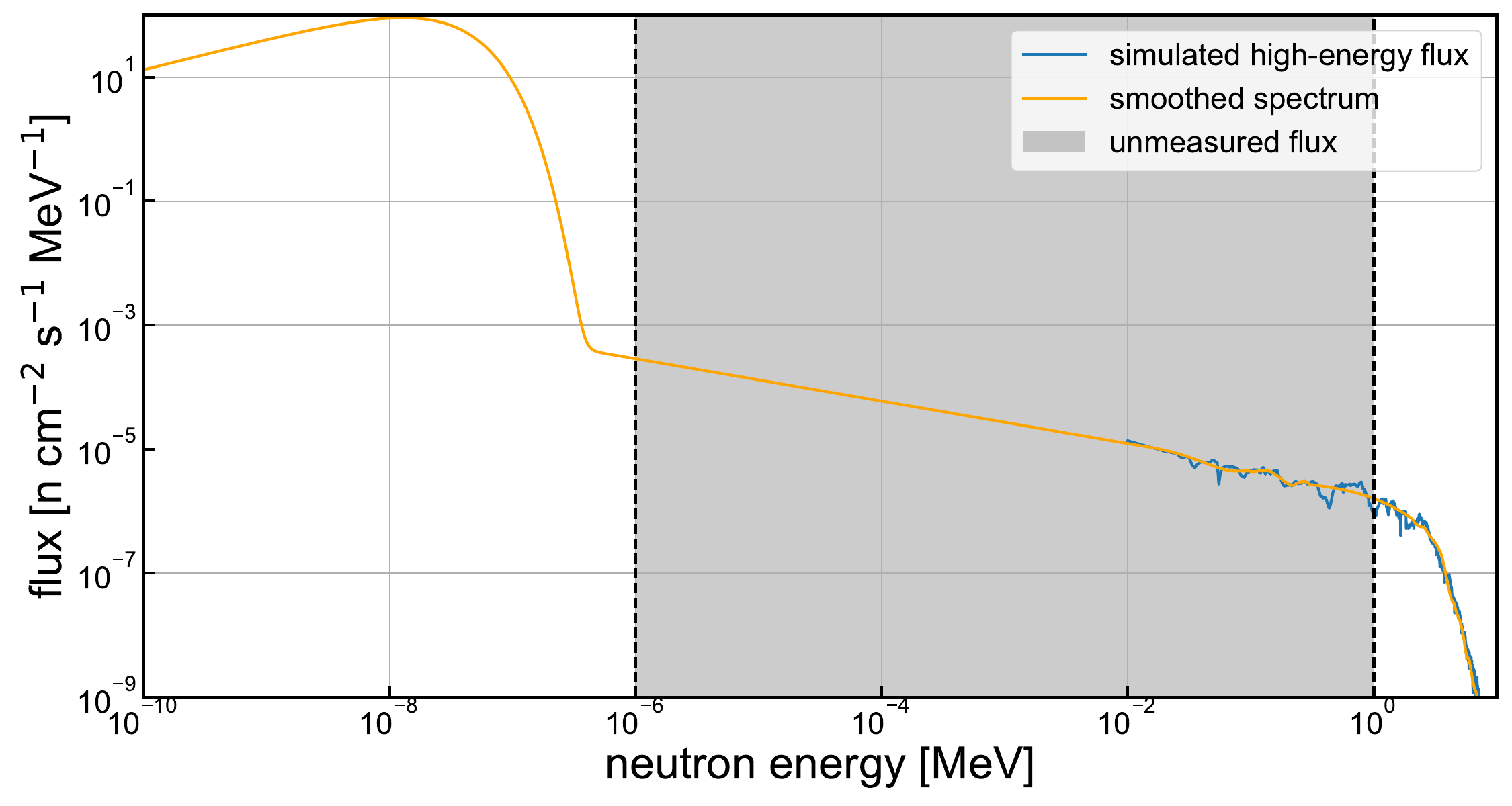}
    \caption[SNOLAB Flux]{Neutron flux as a function of energy for the SNOLAB environment.
at high energy the curve is a smoothing of the high-energy flux (blue) computed from the SuperCDMS
sensitivity projection paper~\cite{PhysRevD.95.082002}. The middle portion of the curve with the
grey shading is an
interpolation in the region below 1\,MeV down to the Maxwell-Boltzmann portion for the thermal
neutron flux. The integral of the thermal flux region (from 10$^{-4}$\,eV to about 1\,eV) is
normalized to the measured underground SNOLAB thermal neutron flux: 4144.8$\pm$49.8$\pm$105.3
n/m$^2$/day.  
}
    \label{fig:flux_function}
\end{figure}

Figure~\ref{fig:flux_function} shows the neutron flux as a function of energy at SNOLAB. It was
estimated from a combination of studies done in the past. First was a Monte Carlo approach
targeting radiogenic neutrons from the cavern walls~\cite{PhysRevD.95.082002}, this constrained
the shape of the distribution for 6--10\,MeV neutrons. The overall normalization for the ``fast''
neutrons was taken from the SNOLAB handbook~\cite{SNOLABhandbook}--the value was
4000\,n/m$^2$/day. The thermal flux--measured for SNO--is taken to come from a
Maxwell-Boltzman spectrum with overall normalization that comes from measurement. In this case the
measurement is that of Browne~\cite{Browne:1999pe} and is quoted with the mean value first
followed by the statistical and systematic uncertainties respectively: 4144.9$\pm$49.8$\pm$105.3
n/m$^2$/day. The intervals on each of the types of uncertainty were assumed Gaussian and
symmetric. The flux in the area between the thermal and high-energy measurements is generally
unmeasured and has been interpolated with a simple power-law function in
Fig.~\ref{fig:flux_function}, $aE^b$. Our prototype detector will measure the flux in this energy
region which is especially important for low-mass dark matter searches.

\begin{figure}[h!]
    \centering
    \includegraphics[width=0.6\linewidth]{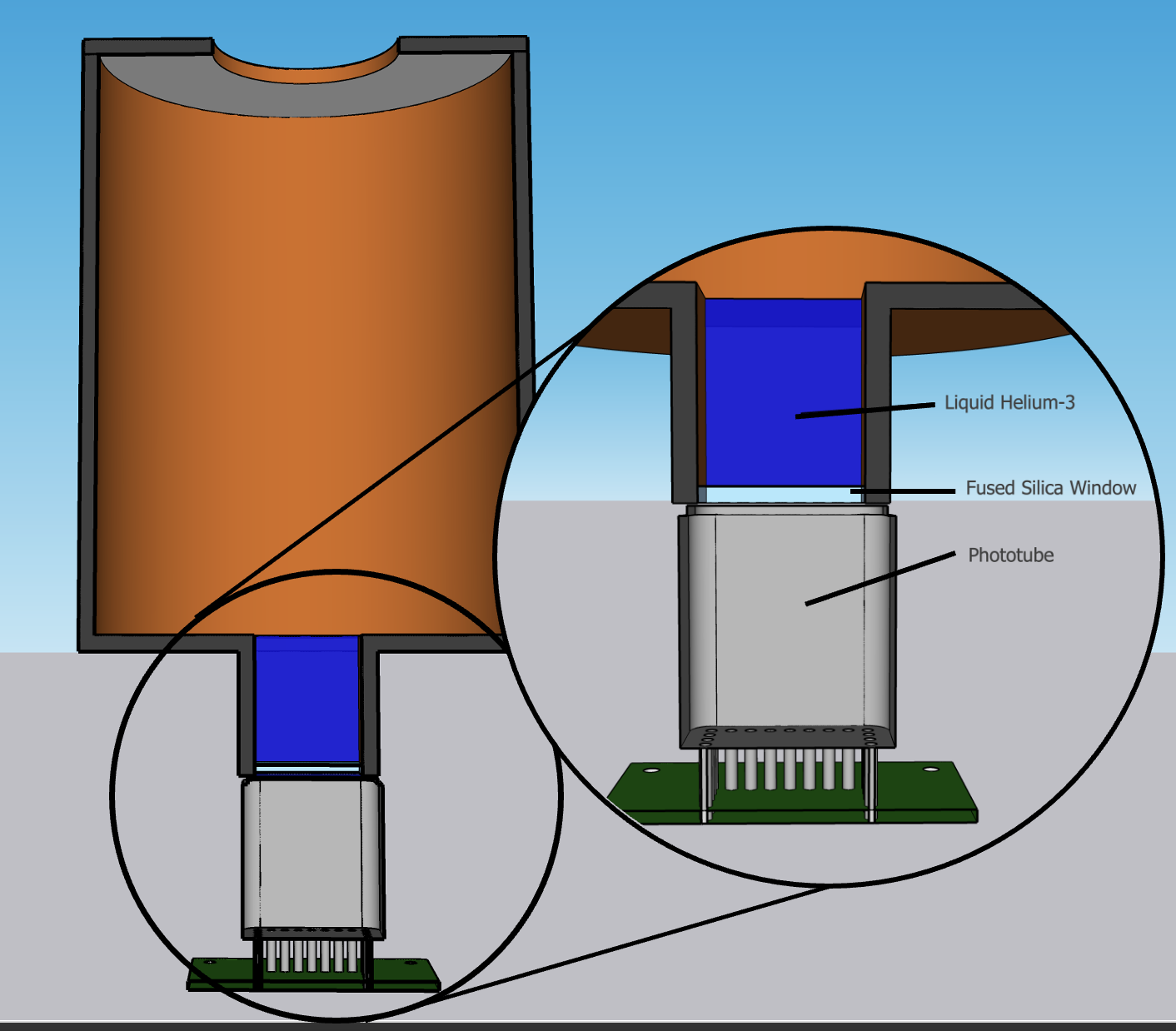}
    \caption[Detector Drawing]{The prototype design of a liquid $^3$He neutron detector with a
photomultiplier readout. The height of the gaseous portion of the vessel is approximately 11\,cm
tall, whereas the liquid fills a 2$\times$2$\times$2\,cm$^3$ cubical vessel attached near the bottom.
}
    \label{fig:detector-drawing}
\end{figure}

Recently the SPICE/HeRALD collaboration provided a proof-of-principle for a cryogenic
photomultiplier readout of a liquid helium volume~\cite{PhysRevD.105.092005}. The collaboration
also kindly shared their published data with us in digital form which we use for our design.
Other collaborations like QUEST-DMC, and MACHe3 have also proposed the use of liquid $^3$He directly
for dark matter searches~\cite{Autti2024,autti2023long,MAYET2000554}. Our basic design is shown in
Fig.~\ref{fig:detector-drawing}. We have envisioned a hermetic copper vessel with a large gas
space above a small cube (2$\times$2$\times$2\,cm$^3$) where liquid $^3$He will collect when the
vessel is filled with 6.63\,liters (22.9\,atm) of $^3$He at room temperature and lowered to a
temperature of 1\,K. Liquification will be achieved by attaching the vessel to an off-the-shelf,
commercially obtainable closed-cycle 1\,K cryocooler~\cite{cryocooler}. The cubical portion of the
vessel will have a small quartz window through coated with tetraphenyl butadiene (TPB) so the
phototube can measure the helium scintillation~\cite{MCKINSEY1997351}.

\section{\label{sec:detperformance}Expected Spectrum}

\begin{figure}[h!]
    \centering
    \includegraphics[width=1.0\linewidth]{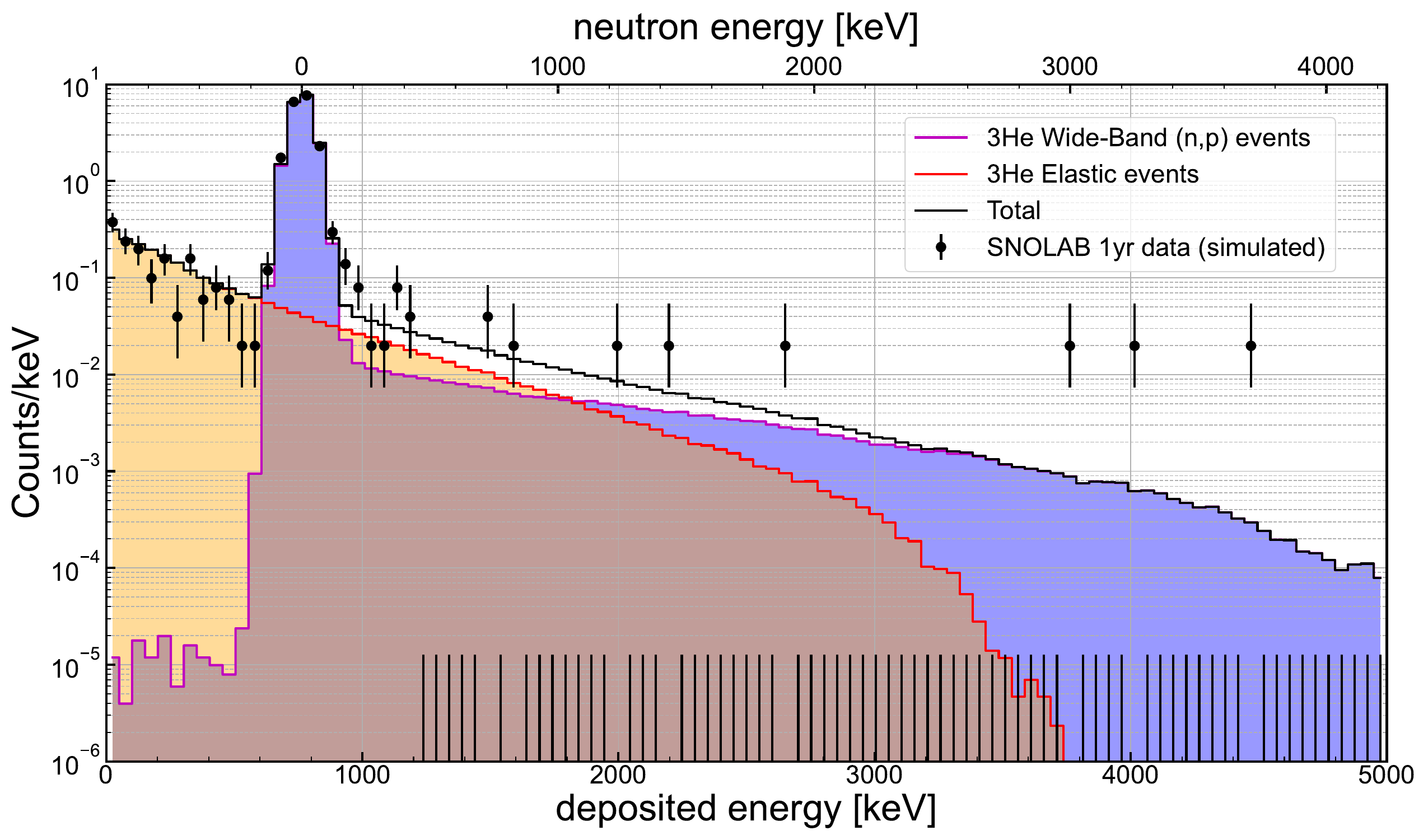}
    \caption[Expected Spectrum]{The expected spectrum of a 2$\times$2$\times$2\,cm$^3$ detector
operated at SNOLAB for 1\,yr (black points) compared to the expected total neutron distribution
(black line) with (n,p) (blue shaded) and elastic (orange shaded) contributions. The uncertainties
for zero-count bins were scaled so they did not obscure the plot, the points appear as black
spikes on the plot ending at a vertical axis value of 10$^{-5}$.  }
    \label{fig:recoil}
\end{figure}

My research group has estimated the measured neutron spectrum for a 1\,yr SNOLAB exposure of the
2$\times$2$\times$2\,cm$^3$ liquid $^3$He volume~\cite{sharbaugh2023response}. We used the SNOLAB
spectrum of Fig.~\ref{fig:flux_function} as an input spectrum and obtained the deposited-energy
spectrum shown in Fig.~\ref{fig:recoil}

Our liquid $^3$He neutron detector design is appropriate for low neutron flux environments because
of its high efficiency relative to conventional gaseous $^3$He tubes like those used by
Beimer~\cite{BEIMER1986402}. This is because of the much higher density of $^3$He in the liquid
state at 1\,K--0.0792\,g/cm$^3$--compared to a 6\,atm gas with density
7.4$\times$10$^{-4}$\,g/cm$^3$ at room temperature~\cite{PhysRev.96.551}. In fact, our
2$\times$2$\times$2\,cm$^3$ detector is expected register 3 times more neutron events above around
10\,keV than the larger tubes (5\,cm diameter; 15\,cm length) of Beimer~\cite{BEIMER1986402}. For
typical gaseous tubes with fill pressures between 6--10\,atm our design will be 64--107 times more
efficient per volume at neutron energies larger than 10\,keV. Below 10\,keV the efficiency will
quickly approach 100\%.  

Figure~\ref{fig:recoil} shows that for a 1\,yr exposure we will produce an excellent
measurement of the thermal neutron flux--given by the size of the peak near a deposited energy
equal to the Q-value of the (n,p) reaction. It is also likely that a run period of that time will
produce 3--4\,counts in the high-energy neutron region above 2700\,keV. Neutrons of those energies
will constrain the overall fast flux present at the detector location. Similarly, the counts in
the intermediate region, above around 266\,keV neutron energy but below 1000\,keV will create an
important constraint in a region that has previously been difficult to measure.  

\section{\label{sec:bknds}Backgrounds to Neutron Measurement}

When trying to measure low neutron fluxes with $^3$He--in a gaseous or liquid state--the dominant
backgrounds are probably $\alpha$ particles from the boundaries of the active volume. There are
also lesser backgrounds from ambient gamma rays and radiogenic neutrons from the detector
materials. Our liquid $^3$He design allows a large number of $^3$He atoms to be placed in a
container with relatively low surface area compared to gaseous detectors. This generically limits
the dominant background.

A good example is the Helium and Lead Observatory (HALO)--a gaseous $^3$He array operating at
SNOLAB~\cite{Bruulsema:2017}.  All together the array consists of around 0.75\,m$^3$ of $^3$He at
2.5\,atm using special ultra-pure nickel tubes that reach a thorium contamination of
1\,ppt~\cite{BOGER2000172}.  That array will produce about 11,000 neutron counts above 1\,keV a
year, but around 140,000 $\alpha$ background events. Our design with just 8\,cm$^3$ active volume
will only produce 30 neutron events above 1\,keV but could be expected to have around 115 $\alpha$
background events in the same time period if we use highly radiopure copper (see below); a
signal-to-noise ratio over two times better.

Actually, the background of our prototype can be improved substantially with a re-designed
cryogenic readout. The background budget of our design is shown in Table~\ref{tab:bknd_budget}.
Each of the major background-contributing components are listed along with their lowest-available
thorium (Th) contamination in parts per trillion (ppt). By far the largest contributions come from
components related to the photomultiplier readout--the quartz window and TPB coating.

\begin{table}[h!] 
\caption{\label{tab:bknd_budget}Background budget for the liquid $^3$He design using a quartz
window, TPB wavelength shifter, and photomultiplier readout. Other materials that generally make
up cryogenic setups but are not envisioned to be near enough to the detector volume to impact the
signal-to-noise are also listed for comparison. Those measurements are taken from the QUEST-DMC
project~\cite{Autti2024}. The table addresses the alpha backgrounds only.}
\begin{tabular}{cccc} 
\toprule Component & Contamination Level (ppt) & Background ($\alpha$/yr) & Best signal-to-noise\footnotemark[1] \\ 
\midrule 
 standard copper\footnotemark[2]  & 39 & 179 & 0.10 \\ 
ultra clean copper\footnotemark[3]  & 0.03 & 0.138 & 0.26 \\ 
quartz window\footnotemark[4] & 2 & 1.9 & 0.26 \\ 
TPB coating  & 120 & 112 & 0.26 \\ 
\midrule
\multicolumn{4}{c}{other materials not directly contacting sensitive
volume\footnotemark[5]} \\ 
\midrule 
stainless steel  & 959 & 4421 & - \\ 
aluminum  & 8.76$\times$10$^4$ & 4.04$\times$10$^5$ & - \\ 
stycast  & $<$ 3.15$\times$10$^3$ & $<$ 1.45$\times$10$^4$ & - \\ 
\botrule 
\end{tabular} 
\footnotetext[1]{assuming all other components optimized for low background with photomultiplier
readout} 
\footnotetext[2]{copper wire used in the Majorana project~\cite{ABGRALL201622}}
\footnotetext[3]{as demonstrated by the Majorana project~\cite{ABGRALL201622} using electroformed
copper}
\footnotetext[4]{as demonstrated by the EXO project~\cite{LEONARD2008490}}
\footnotetext[5]{measurements take from the QUEST-DMC project~\cite{Autti2024}}
\end{table}

Table~\ref{tab:bknd_budget} suggests to move to a scintillation readout that does not require a
standard phototube or quartz window. This can be done using a bolometric Transition Edge Sensor
(TES) or Microwave Kinetic Inductance Device (MKID) mounted on a silicon substrate \emph{inside}
the hermetic copper volume of our detector. Mounting just above the liquid $^3$He volume will
allow good light collection without a quartz window. Silicon substrates can be made very clean and
will not contribute much to the thorium contamination budget. With this modification we would
expect to have a signal-to-noise ratio for neutrons above 1\,keV of around 217 if we use ultra
clean copper. This advancement would make for a high sensitivity neutron flux measurement even in
very low background environments. Furthermore, the resolution and energy distributions of the
signal and background can be taken into account for further improvements in the measurement.

There are detectors--like those proposed by the QUEST-DMC collaboration--that may realize
the ability to use superfluid $^3$He and implement some sort of heat
readout~\cite{Autti2024,autti2023long}. This readout in addition to the scintillation readout we
propose might allow for particle discrimination--the ratio of scintillation to heat is typically
different for heavier recoils. In that case significant background mitigation might be achievable
in our measurement because we could select proton recoils and exclude electron or alpha recoils.
In fact Winkelmann and collaborators~\cite{WINKELMANN2007264} have shown that for proton and
tritium recoils resulting from the $^3$He(n,p) reaction on superfluid $^3$He will produce less
than 15\% scintillation fraction whereas cosmic-ray muons and low-energy electrons will have
around 21\% and 23\% scintillation fractions respectively.  The possibility of heat measurement
would increase the cost of the prototype explored here but may warrant further study.

\section{\label{sec:ack}Acknowledgments}
I gratefully acknowledge the SPICE/HeRALD collaboration for providing us with the data from
several plots in their paper. I would also like to acknowledge the work of Alya Sharbaugh and Luke
Jones on this topic.

\clearpage
\bibliography{ltd20refs_short}


\end{document}